# Directed Graph-based Wireless EEG Sensor Channel Selection Approach for Cognitive Task Classification

Abduljalil Mohamed*, Khaled Bashir Shaban**, and Amr Mohamed**

*Information Systems Department, Ahmed Bin Mohamed Military College
P.O. Box: 22988, Doha, Qatar

**Computer Science and Engineering Department, College of Engineering, Qatar University
P.O. Box: 2713, Doha, Qatar

ajamoham@abmmc.edu.qa, khaled.shaban@qu.edu.qa, amrm@qu.edu.qa

*Abstract*—**Wireless electroencephalogram (EEG) sensors have been successfully applied in many medical and computer brain interface classifications. A common characteristic of wireless EEG sensors is that they are low powered devices, and hence an efficient usage of sensor energy resources is critical for any practical application. One way of minimizing energy consumption by the EEG sensors is by reducing the number of EEG channels participating in the classification process. For the purpose of classifying EEG signals, we propose a directed acyclic graph (DAG)-based channel selection algorithm. To achieve this objective, the EEG sensor channels are first realized in a complete undirected graph, where each channel is represented by a node. An edge between any two nodes indicates the collaboration between these nodes in identifying the system state; and the significance of this collaboration is quantified by a weight assigned to the edge. The complete graph is then reduced into a directed acyclic graph that encodes the knowledge of the non-increasing order of the channel ranking for each cognitive task. The channel selection algorithm utilizes this directed graph to find a maximum path such that the total weight of this path satisfies a predefined threshold. It has been demonstrated experimentally that channel utilization has been reduced by 50% in the worst case scenario for a three-state system and an EEG sensor with 14 channels; and the best classification accuracy obtained is 81%.**

## I. INTRODUCTION

Recent advances in wireless communication and intelligent wireless electroencephalogram (EEG) sensors have allowed the realization of different EEG-based applications, such as healthcare [1] and brain computer interface applications [2]. In general, an EEG sensor network topology comprises of simple sensors that collect information about the subject and send it through wireless paths to the sink. A common characteristic of the wireless EEG sensors is that they are low powered devices, and hence an efficient usage of sensor energy resources is critical for any practical application. As shown in [3, 4], the sensing and processing energy are negligible with respect to communication energy. Thus, most of the energy-aware algorithms reported in the literature address this issue at the communication level [5-7]. It is also shown in [8] that the reduction of the number of sensors can also reduce the power consumption for wireless EEG caps. Moreover, it yields more comfort for the user, decreases installation time duration and may substantially reduce the financial cost of the Brain Computer Interface (BCI) setup since the cost of an EEG cap and an amplifier vary in relation to the number of channels [9].

In this paper we address the problem of channel reduction for multi-channel sensor acquiring EEG data for a BCI classifier identifying cognitive tasks. The channel selection problem can be well represented as a complete undirected graph, where each channel is represented as a graph node. An edge between any two nodes indicates the collaboration between the corresponding channels in identifying the current system state; and the significance of this collaboration is quantified by a weight assigned to the edge.

Various approaches that tackle the issue of utilizing only an informative subset of channels rather than the complete set of the available sensor channels have been presented recently in the literature. In [9], a sensor channel selection method based on the backward elimination is proposed. The method uses a cost function that is based on the signal to signal-plus-noise ratio with spatial filtering. The evaluation of the selected subset of channels is assessed on three different levels: 1) at a global level, the measure of the signal EEG to noise, 2) at a recognition level, the overall accuracy of the P300 event detection, 3) the accuracy of the speller application. The conducted experiments have shown that selection methods do not consider the spatial filters provide the worst results. In [10], the authors use phase locking values to measure the variability of phase difference between two EEG signals. It basically characterizes the behavioral similarity between two channels. The performance of the selected subset of channels is evaluated by measuring the character recognition accuracy. Thomas N. L. et al [11], propose a Support Vector Machine-based channel selection method. Without using prior knowledge about the spatial distribution of brain activity of a mental task, the method utilizes the recursive feature elimination (RFE) and zero-norm optimization (10-Opt) based on support vector machine. All features of one channel are either removed completely (the channel is removed) or all features remain. The proposed methods are applied to motor imagery and have shown that, while REF is capable of reducing the channels needed for a robust classification without an increase of error, the performance of the 10-Opt algorithm is less than satisfactory. In [12], the authors propose a multichannel classification method that uses a committee of artificial neural networks (ANNs). The coefficients of the power spectrum estimator autoregressive (AR) model are used as feature vectors representing the data in one channel. For each individual channel in the EEG, an ANN is trained to classify three classes (left and right index fingers, and right foot). The classification accuracy depends on the number of the committee members (channels), however, the average classification success does not vary significantly with committee members with sizes bigger than 20 channels. In [13], a synchronization likelihood-based approach is proposed. It basically measures both the linear and nonlinear interdependency between signals. The synchronization likelihood (SL), along with Hjorth parameters, are used as input features to the Linear Discriminant Analysis classifier. The SL was computed in a sliding window of size 512 sample points. Taking into account both anatomical knowledge and SL values, five channels from different brain regions have been selected among the EEG channels which exceed the global mean of SL value. Utilizing these five channels, using different features, the LDA classifier yields a better classification accuracy than it does utilizing all the EEG channels. In [14], the EEG channels are ranked such that the mutual information (MI) between selected sensors and class labels are maximized. An efficient computational approach to estimate the MI is presented. In [15], the principal component analysis is used as a feature selection method. The selected channels are to preserve as much information as compared to the complete set of channels as possible. A multilayer perceptron (MLP) neural network is used to classify the EEG data into two categories: alcoholics and non-alcoholics. Experiments show that, on average, classification performance using the full set (64 channels) is not significantly different from using 16 channels. However, as the number of channels is reduced to 8 and 4, they produce a drastic drop in the classification accuracy.

The rest of the paper is organized as follows: Section II introduces the channel selection representation model. The basic notions of the evidence theory are introduced in section III. The proposed approach is detailed in section IV. Section V reports the results obtained by the new approach and final remarks are summarized in section VI.

## II. CHANNEL SLECTION REPRESNTATION MODEL

The sensor channel selection representation model plays an essential role in the new proposed algorithm. Basic graph concepts and notations are first introduced in this section to facilitate the construction of the graph-based proposed algorithm.

### A. Basic Notations

We consider an EEG sensor consisting of N channels. We represent the sensor channels with an undirected weighed graph $G = (V, E)$, where $V = \{v_1, v_2, ..., v_n\}$ is the set of nodes (channels), and $E = \{(v_i, v_j) \mid f(x_{vi}, x_{vj}) > 0\}$ is the set of edges, and f(.) is the task weight function (or weight function for simplicity) and defined as follows:

$$f : (x_{vi}, x_{vj}) \rightarrow (0,1]$$

where $x_{vi}$ and $x_{vj}$ are the current readings of the channels $v_i$ and $v_j$, respectively. We assume the existence of this weight function. The edge between any two graph nodes sensor channels does not mean a communication link, but rather it indicates the collaboration between these two channels in identifying the current system state. Using certain measures, the weight function assigns to this edge a positive value that reflects the confidence of the collaborating channels in their decision. Moreover, it assigns a confidence value for each individual channel node. Hence, the weighting function plays a crucial role and should be contrived carefully.

### B. General Graph-based Channel Selection

Given a representation graph model, the problem now is to find a path that starts from a certain channel node and monotonically increases, in terms of confidence value, as an additional node is added to the path. Fig.1 demonstrates the channel search process for an EEG sensor with six channels. It starts with the channel node (node 3) that has the maximum confidence value. It then searches for the next node such that its collaboration with previous node yields the highest confidence value among all other nodes. The directed edge in the figure refers to the next selected channel node. With the exception of the edge connecting the selected node with its previous one, all the edges of the selected node are removed from the graph. The search process continues in this manner until either the accumulated path confidence value exceeds the defined threshold or all the available channels are visited.

## III. DIRECTED ACYCLIC GRAPH-BASED CHANNEL SELECTION ALGORITHM

The computational complexity associated with the *general graph-based channel selection* is due to the exhaustive search resulted from the undirected complete graph model. Without loss of generality, we will assume a directed acyclic graph model in which the new search method is performed in a systematic and more efficient way. The new channel selection algorithm always starts from a special node (or nodes) which will be denoted as the leading node. The structure of the directed edges in the new graph model guides the selection process to the next most appropriate node based on the path confidence value assigned to the edges. The search process stops when it reaches the destination node (a dummy node that is designated as the final node). Clearly, the graph topology

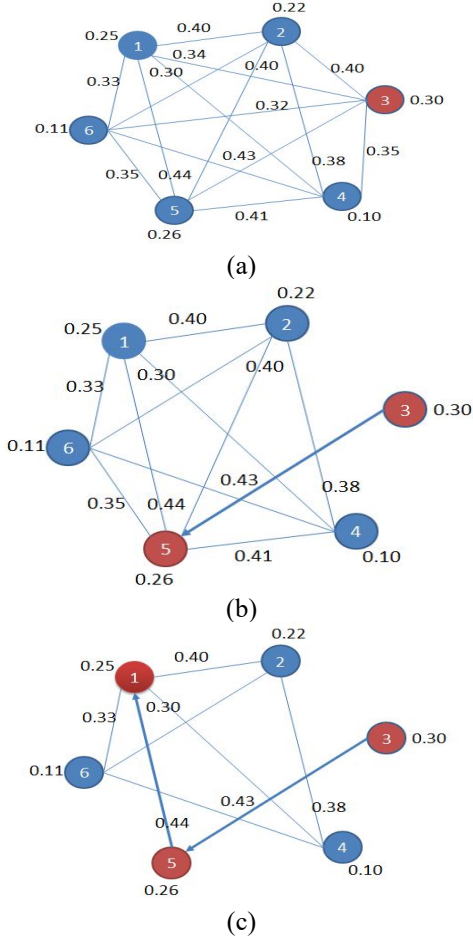

Figure 1. General graph-based channel selection algorithm (a) node 3 is selected as a starting channel (b) node 5 is selected next and added to the search path (c) node node is added as a final node, assuming the total confidence value exceeds a predefined threshold

plays a critical role in the search process. Domain knowledge can be used to construct such topology. In this work, the EEG sensors are used for the purpose of cognitive task classification. Thus, as a first step in the graph construction process, the sensor channels are ranked based on classification measures.

*A. Sensor Channel Ranking*

The most-widely sensor channel ranking method reported in the literature is based on the overall performance of the channel selection criterion. It basically utilizes the global knowledge as it effectively identifies the most relevant sensors. An irrelevant sensor is a sensor whose removal barely impairs the performance of the selection criterion. However, this method ignores the local knowledge that the sensors may have. Experimental work has shown that different sensors, though they may exhibit similar overall performance, may actually have different responses to different cognitive tasks. Hence, the relativity of a certain channel to a given task is a more appropriate ranking measure. For this purpose, the precision measure is used. Using an appropriate classifier model, the overall classification accuracy of each sensor channel and its sensitivity to each task hypothesis can be computed. The precision measure can be given a probabilistic interpretation as the probability that a random channel reading is relevant to the given cognitive task. Based on the above discussion, the following definitions will be used in the graph construction process.

**Definition 1** (channel ranking set). *Let $P_{wk}^{vi}$ refer to the class precision of the channel node $v_i$ for the task $w_k$. The channel ranking set, $C^{wk}$, for the task $w_k$ is defined as follow:*

$$C^{wk} = \{v'_1, v'_2, ..., v'_N\},$$

$C^{wk}$ *is an ordered permutation of the set $V$ such that the following condition holds:*

$$P_{w_k}^{v_1} \geq P_{w_k}^{v_2} \geq ... \geq P_{w_k}^{v_N},$$

**Definition 2** (leading channel). *Given the channel ranking set, $C^{wk}$ as defined in Definition 1, the channel node $v'_1 \in C^{wk}$ is called the first leading channel for the task $w_k$, $v'_2 \in C^{wk}$ the second leading channel for the task $w_k$, and so on.*

*B. Constructing Directed graph*

Given channel ranking sets, as defined in Definition 1, a directed graph can be constructed as follows: the first leading channels of each ranking set constitute the first layer of the directed graph, the second leading channels constitute the second layer, and so on. Identical leading channels are replaced with their successive channels in their respective ranking sets. Directed edges are added from every node in a each layer to every node in the subsequent layer. The nodes of the last layer are all connected to a final destination node (dummy node), and the nodes of the first layer are merged to constitute the root node of the directed graph. The graph construction process can be better explained by the following example attained from previous work [16]. The 14-channel Emotiv EPOC EEG sensor was used for the purpose of classification of three-mental tasks (T1, T2, T3).

Using the RapidMiner software tool [17], the quality of the discriminant power of each channel is examined. The *k*-NN classifier model is utilized, where the *k* parameter is set to 4. The channel classification results are shown in Table I. In the table, the first fourteen columns refer to the fourteen Emotiv EEG channels numbered from 4 till 17, respectively. The tasks T1, T2, and T3 correspond to the three mental tasks; sending an email, dialing a phone number, and initiating a Skype session, respectively, and indicate the class precision obtained by the classifier. The

TABLE 1 k-NN BASED CHANNEL RANKING

| Channels | | | | | | | | | | | | | | Tasks | | | Accuracy |
|---|---|---|---|---|---|---|---|---|---|---|---|---|---|---|---|---|---|
| 4 | 5 | 6 | 7 | 8 | 9 | 10 | 11 | 12 | 13 | 14 | 15 | 16 | 17 | T1 | T2 | T3 | Overall |
| √ | X | X | X | X | X | X | X | X | X | X | X | X | X | 33.77% | 51.83% | 48.13% | 34.50 |
| X | √ | X | X | X | X | X | X | X | X | X | X | X | X | 33.21% | 37.98% | 40.23% | 33.33% |
| X | X | √ | X | X | X | X | X | X | X | X | X | X | X | 33.48% | 45.95% | 82.38% | 33.90% |
| X | X | X | √ | X | X | X | X | X | X | X | X | X | X | 33.64% | 48.67% | 72.11% | 34.19% |
| X | X | X | X | √ | X | X | X | X | X | X | X | X | X | 33.56% | 34.69% | 73.88% | 34.05% |
| X | X | X | X | X | √ | X | X | X | X | X | X | X | X | 33.40% | 47.44% | 68.42% | 33.72% |
| X | X | X | X | X | X | √ | X | X | X | X | X | X | X | 33.45% | 57.53% | 73.66% | 33.85% |
| X | X | X | X | X | X | X | √ | X | X | X | X | X | X | 33.60% | 56.57% | **83.54%** | 34.44% |
| X | X | X | X | X | X | X | X | √ | X | X | X | X | X | 33.61% | 54.57% | 69.66% | 34.24% |
| X | X | X | X | X | X | X | X | X | √ | X | X | X | X | 33.97% | **74.42%** | 75.58% | 35.47% |
| X | X | X | X | X | X | X | X | X | X | √ | X | X | X | 33.47% | 41.11% | 64.96% | 33.83% |
| X | X | X | X | X | X | X | X | X | X | X | √ | X | X | **34.56%** | 75.31% | 48.25% | **36.31%** |
| X | X | X | X | X | X | X | X | X | X | X | X | √ | X | 34.00% | 58.13% | 57.11% | 34.94% |
| X | X | X | X | X | X | X | X | X | X | X | X | X | √ | 33.63% | 43.52% | 67.17% | 34.17% |

last column is the overall classification accuracy for each channel. The symbol √ means the respective channel is used for the EEG signal classification, while the symbol X denotes its absence. Notably, no single channel yields a satisfactory performance level in terms of the overall classification accuracy (less than 37% accuracy). However, it can be noted from the table that EEG channels respond differently to different mental tasks. For T1, channel 15 performs best at 34.56% class precision rate. For T2, channel 13 outperforms other channels at 74.42% rate. For T3, channel 11 yields the best class precision rate at 83.53%. The channels 11, 13, and 15 constitute the initial observed channels. Hence, the root node, $v^*$, is:

$$v^* = \{11,13,15\},$$

Based on the classification results shown in Table I, the channel ranking sets are determined as follows:

$$C^{T1} = \{16,7,4,17,12,8,14,6,10,9,5\},$$
$$C^{T2} = \{16,10,12,4,7,9,6,17,14,5,8\},$$
$$C^{T3} = \{6,8,10,7,12,9,14,17,16,4,5\},$$

The obtained directed acyclic graph is shown in Fig. 2. In the graph, the $R$ node represents the root node which refers to the leading nodes for each mental task and consists of the nodes 11, 13, and 15. The last node in the graph, $D$, is the destination node, which serves as a flag indicating the end of the channel search process.

### C. Directed Acyclic Graph based Cahnnel Selection Algorithm

A new directed acyclic graph-based channel selection (*DBCS*) algorithm can now be developed. Let $\overline{G} = (\overline{V}, \overline{E}, v^*, v_d)$ be a directed acyclic graph of size $N$, constructed as shown in the previous section, such that $\overline{V}$ is the set of nodes, $\overline{E}$ is the set of edges, $v^*$ is the root node, and $v_d$ is the destination node. The *DBCS* algorithm is formally presented by the following pseudo-code.

## IV. PERFORMANCE EVALUATION

In this section, the performance of the proposed channel selection *DBCSA* approach is examined.

**Algorithm**. (DAG-based channel selection (DBCS))

Let $v_x$ be an auxiliary node,

Let $\overline{e}_{ij}$ be a directed edge from the node $v_i$ to the node $v_j$,

Let $c\overline{e}_{ij}$ be the confidence value of the edge $\overline{e}_{ij}$,

Let $\overline{E}_{vi}$ be the set of all the outgoing edges of the channel $v_i$, $\overline{E}_{vi} \in \overline{E}$,

Initializations:
$$PN = \{\phi\},$$
$$CR = \{\phi\},$$
$$PCV = 0$$

1. For all $\overline{e}_{ij} \in \overline{E}$,
   calculate $c\overline{e}_{ij} = f(x_{vi}, x_{vj})$,
2. $v_x \leftarrow v^*$,
3. Add the nodes of the set $v_x$ and their current readings to the sets $PN$ and $CR$, respectively:
   - $PN \leftarrow \{PN + \{v_x\}\}$
   - $CR \leftarrow \{CR + \{x_{vx}\}\}$
4. Calculate the $PCV = f(CR)$,
5. If $PCV \geq Threshold$, Go To 10,
6. Select an edge $\overline{e}_{xj} \in \overline{E}_{vx}$ such that $c\overline{e}_{xj}$ is a maximum,
7. Set the head node of the edge $\overline{e}_{xj}$ to the node $v_x$, $v_x \leftarrow neighbor(\overline{e}_{xj})$,
8. If ($v_x == v_d$), Go To 10
9. Go To 3,
10. Report $PN$,

END

### A. Data Collection

Using a 14-channel EEG sensor, signals were collected from five male subjects, referred to as S1, S2, S3, S4, and S5. The data collection procedure outlined in [16] was used. Each subject was asked to perform three mental tasks namely, sending an email, dialing a phone number, and initiating Skype session. Interested readers should refer to [16] for more details.

### B. AAN and DBCS Cognitive Classification Accuracy

Since cognitive task identification is considered a classification problem, ANN with scaled conjugate gradient

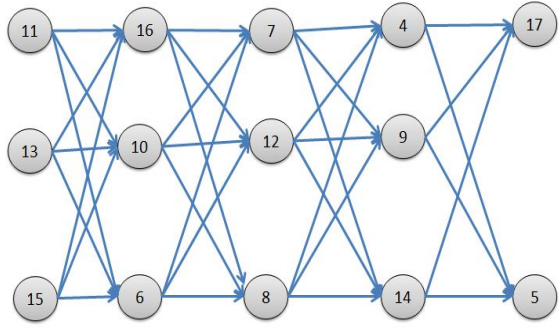

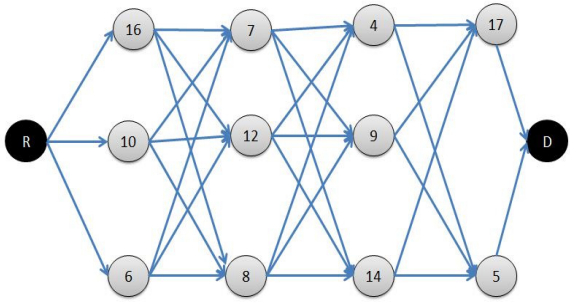

Figure 2. Directed graph-based representation model of the Emotive EEG sensor with 14 channels (a) channels are connected by directed edges based on their task sensitivies (b) nodes of the first layer are merged into one node (R), and nodes of the last layer are connected to a final dummy node (D)

back-propagation learning algorithm is used as a classifier. For each subject, 13438 samples were collected. The data were divided into three parts as follows: 70% used for training, 15% used for validation, 15% used for testing. The purpose is to examine the contribution of the complete set of the sensor channels to the classification task. Employing the 14 channels, the MLP network classification accuracies of the five subjects are shown in Table II. As shown in the table, certain subjects are more susceptible to particular tasks than others. For example, S2 yields the best performance at 95.4% overall accuracy, and S5 yields comparatively lower performance at 91.4% overall accuracy. However, while S2 shows better accuracy than S2 for tasks 2 (98.6%) and S3 (95.7%), S5 performs better in identification task 1 (93.7%) than S2. Similar observations can be made for other subjects.

Moreover, ANNs are also used to learn and estimate the weight values for all the edges and paths of the constructed directed graphs. For the DBCS algorithm, empirically, a threshold value of 0.5 is found to be appropriate. A larger threshold value would require a larger number of channels than the upper bound, and thus a smaller and sufficient number of channels may not be obtained. A smaller threshold can lead to a better channel utilization, however, the cognitive task classification accuracy may degrade to an unacceptable level. The performance of DBCS in terms of classification accuracy is shown in Table III. It can be seen from the table that the best performance is obtained for subject S2 with 81% classification accuracy, whereas the worst performance is obtained for subject S3 with 72% classification accuracy. As expected, the classification accuracy of the proposed algorithm would be lower, but acceptable, than that of the ANN classifier shown in Table II, as the latter uses all the sensor channels during the classification process.

However, the proposed algorithm is superior in terms of channel utilization. On average, the DBCS uses only 20% of sensor channels during the classification process. The distribution of channel utilization for all subjects is shown in Figs. 3. It can be summarized as follows: the root node (lower bound), is used 17% for S2, 15% for S2, and 16% for the subjects S3, S4, and S5. The root node is the lower bound on the channels used. The maximum number of channels (upper bound) used is seven channels and utilized as follows: 23% for S1, 25% for S2, 24% for S3, 25% for S4, and 26% for S5. Due to its adaptive property, the number of channels selected by the DBCS algorithm varies from one subject to another.

## I. CONCLUSION

In this paper, a directed graph-based wireless EEG sensor channel selection approach for cognitive task classification is proposed. The channel selection problem is modeled within the framework of graph theory, where graph nodes represent sensor channels, and graph edges represent the collaboration between the corresponding sensor channels in the classification process. The sensor channels are first ordered based on their task sensitivities. A directed graph is then constructed that encodes the node ordering knowledge into the graph topology. Using the constructed graph, the proposed approach selects the most appropriate channels to identify the current cognition task. For an EEG sensor with 14 channels, the proposed approach utilizes only 7 channels (i.e., 50% channel utilization), in the worst case scenario, and it yields classification accuracy at 81%. In future work, feature selection component can be incorporated within the proposed approach to enhance the model performance in terms of classification accuracy.

TABLE 2 THE NEURAL NETWORK CLASSIFICATION ACCURACY OF THE FIVE SUBJECTS

| Subjects | Classification Accuracy | | | |
|---|---|---|---|---|
| | Task 1 | Task 2 | Task 3 | Overall |
| S1 | 92.8% | 91.5% | 92.4% | 92.2% |
| S2 | 92.2% | 98.6% | 95.7% | 95.4% |
| S3 | 93.3% | 95.7% | 92.5% | 93.8% |
| S4 | 95.4% | 91.7% | 92.6% | 93.2% |
| S5 | 93.7% | 88.4% | 92.4% | 91.4% |

TABLE 3 OVERALL CLASSIFICATION ACCURACY OF THE DBCS ALGORITHM

| Subjects | Classification Accuracy | | | |
|---|---|---|---|---|
| | Task 1 | Task 2 | Task 3 | Overall |
| S1 | 87.66% | 75.70% | 74.46% | 78.12% |
| S2 | 77.34% | 88.44% | 76.88% | 81% |
| S3 | 70.40% | 77.38% | 68.18% | 72% |
| S4 | 78.68% | 79.08% | 78.72% | 79% |
| S5 | 71.52% | 74.72% | 75.08% | 73% |

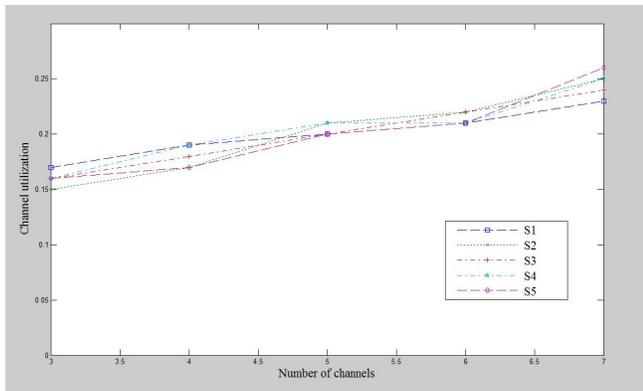

Figure 3. Channel utilization for all subjects

## II. ACKNOWLEDGMENT